\documentstyle[12pt,aasms]{article}
\tightenlines

\begin{document}

ApJ Letters, in press

\vskip 24pt

\title
{A New 3.25 Micron Absorption Feature toward Mon R2/IRS-3}

\vskip 48pt

\author
{K. Sellgren}

\vskip 12pt

\affil
{Department of Astronomy, Ohio State University,
174 West 18th Av., Columbus, OH  43210  USA}

\affil
{e-mail: sellgren@payne.mps.ohio-state.edu}

\vskip 24pt

\author
{T. Y. Brooke}

\vskip 12pt

\affil
{Jet Propulsion Laboratory, M/S 169--237, 4800 Oak Grove Dr.,
Pasadena, CA  91109  USA}

\affil
{e-mail: tyb@scn5.jpl.nasa.gov}

\vskip 24pt

\author
{R. G. Smith}

\vskip 12pt

\affil
{Department of Physics, University College, University of New South Wales,
Australian Defence Force Academy, Canberra ACT 2600, Australia}

\affil
{e-mail: rgs@phadfa.ph.adfa.oz.au}

\vskip 24pt

\author
{T. R. Geballe}

\vskip 12pt

\affil
{Joint Astronomy Centre, 660 N. Aohoku Place, University Park,
Hilo, HI  96720 }

\affil
{e-mail: tom@jach.hawaii.edu}

\clearpage

\centerline
{\bf Abstract}
A new 3.2--3.5~$\mu$m spectrum of the protostar Mon~R2/IRS-3 confirms
our previous tentative detection of a new absorption feature near 3.25
$\mu$m.
The feature in our new spectrum has a central wavelength
of 3.256 $\mu$m
(3071 cm$^{-1}$) and has a full-width at half maximum of 0.079 $\mu$m
(75 cm$^{-1}$).
We explore a possible identification with
aromatic hydrocarbons at low temperatures,
which absorb at a similar wavelength.
If the feature is due to aromatics, the derived column density of C--H
bonds is $\sim$1.8 $\times$ $10^{18}$ cm$^{-2}$.  If the absorbing
aromatic molecules are
of roughly the same size as those responsible for aromatic emission
features in the interstellar medium, then we estimate that $\sim$9\%
of the cosmic abundance of carbon along this line of sight would be in
aromatic hydrocarbons,
in agreement with abundance estimates from emission features.

\vskip 24pt


\clearpage

\section
{Introduction}

\vskip 12pt

The C--H stretch absorptions of many of the organic molecules expected
to be formed or condensed on molecular cloud dust lie in the 3.2--3.6 $\mu$m
region, on the long wavelength side of the 3.1 $\mu$m H$_2$O ice band
which dominates the spectrum of embedded sources.
Sellgren, Smith, \& Brooke (1994) recently reported
a tentative detection of a new
absorption feature at 3.25 $\mu$m (3078 cm$^{-1}$) toward Mon
R2/IRS-3, a protostar in the Mon~R2 star formation region
(Beckwith et al. 1976).  Their spectrum had a resolution
$\lambda/\Delta\lambda \approx 720$ at 3.25 $\mu$m.
Here, we present a new spectra of Mon~R2/IRS-3
with a resolution of 1000 which confirms the presence of a 3.25 $\mu$m
feature. Some possible identifications are discussed.

\vskip 24pt

\section
{Observations}

\vskip 12pt

The latest observations of Mon R2/IRS-3 were made on 1994 October
8 at the United Kingdom Infrared Telescope (UKIRT) on Mauna Kea.
The CGS4 long-slit spectrometer
(Mountain et al. 1990)
was used with the 75 lines mm$^{-1}$
grating in first order and the 300-mm focal length camera.  This provided a
wavelength resolution of 0.0033~$\mu$m ($\lambda/\Delta\lambda$
= 1000 at 3.25~$\mu$m).  The spectrometer is designed to have
only one resolution element per pixel, so improved sampling of the
spectrum was obtained by moving the detector by one-third of a
resolution element between individual spectra and repeating this until
two resolution elements were observed by each pixel. The observations
consist of two overlapping grating positions, at 3.16--3.37 $\mu$m and
3.34--3.55 $\mu$m.  The pixel size was 1.55$''$.  The spectrometer
slit was 90$''$ $\times$ 1.55$''$ with the long direction oriented
east-west.  The sources were nodded $\sim$12$''$ along the
slit for background subtraction.  An argon spectrum in second order
was used for wavelength calibration.
We compared our spectrum of Mon R2/IRS-3 with the star HR 1948 (O9Iab:)
for atmospheric cancellation.  The airmass difference between Mon
R2/IRS-3 and HR 1948 was always less than 0.03.

In the final spectra, several points at 3.313 -- 3.321 $\mu$m affected
by strong telluric CH$_4$ have been removed.  We have also removed
points near 3.297~$\mu$m which may have been affected by any
photospheric Pfund $\delta$ feature in the O9Iab: atmospheric
comparison star.

\vskip 24pt

\newpage
\section
{Results}

\vskip 12pt

The new spectrum of Mon R2/IRS-3 is shown in Figure 1.  The
observations fall in the region of the 3.1 $\mu$m H$_2$O ice band and
the broad absorption wing which peaks near 3.3--3.4~$\mu$m
(Smith, Sellgren, \& Tokunaga 1989).  The intrinsic spectral shape of this
absorption is uncertain.  Thus the best continuum to use for deriving
the optical depth of narrow absorption features in this region is a
local continuum which passes smoothly through those parts of the
spectrum not containing narrow absorption features.  We have fit a
second-order polynomial to the spectrum of Mon R2/IRS-3, excluding data
at 3.2--3.3~$\mu$m and longward of 3.4 $\mu$m from the fit.  The choice of
excluded regions is the same as that used by Sellgren et al. (1994).
Our adopted continuum is shown as a solid
line in Figure 1.

The derived optical depth is also shown in Figure 1.  We fit two Gaussians
to the optical depth curve.
The central wavelength, full width at
half-maximum (FWHM), and optical depth of each Gaussian were varied
to produce the best fit to our observations.
We derive central wavelengths of 3.256 $\pm$ 0.003 $\mu$m
and 3.484 $\pm$ 0.003 $\mu$m
(3071 $\pm$ 3 cm$^{-1}$
and 2870 $\pm$ 2 cm$^{-1}$)
for the 3.25 $\mu$m and
3.48 $\mu$m features, respectively.
We also find FWHM values
of 0.079 $\pm$ 0.007 $\mu$m
and 0.117 $\pm$ 0.007 $\mu$m
(75 $\pm$ 6~cm$^{-1}$
and 97 $\pm$ 6~cm$^{-1}$)
for the 3.25 $\mu$m
and 3.48 $\mu$m features, respectively.
Our new measurements of the central wavelengths and widths
agree well with those of Sellgren et al. (1994).
The 3.25 $\mu$m optical depth we measure, 0.045,
also agrees well with Sellgren et al. (1994).
The 3.48 $\mu$m optical depth we derive, 0.058, does not agree
with the value of 0.036 measured by Sellgren et al. (1994).
However,
the optical depth is sensitive to the choice of continuum,
so the Sellgren et al. (1994) spectrum provides the most reliable
value for the 3.48 $\mu$m optical depth because
the current spectrum (Fig. 1) does not extend to long enough wavelengths
to provide continuum on the long wavelength side of the 3.48 $\mu$m
feature.

\vskip 24pt

\section
{Discussion}

\vskip 12pt

The 3.48 $\mu$m feature was first identified by
Allamandola et al.  (1992) toward four protostars.
They attributed the feature to C--H
bonds in hydrocarbons with ``diamond-like'' bonding.
This feature in Mon R2/IRS-3 and other
sources is discussed in more detail by Brooke, Sellgren, \& Smith (1995).

Standard references on room temperature infrared spectra suggest that the
3.25 $\mu$m feature might be due to a C--H stretch of
the =CH$_2$ group in an alkene, which occurs at 3.23--3.25 $\mu$m
(e.g. Williams \& Fleming 1987).
An alkene identification, however, is unlikely because alkenes have a
second, comparably strong, feature at 3.29--3.32 $\mu$m which is not
observed toward Mon R2/IRS-3.
We have searched the low temperature laboratory spectra of
pure ices and ice mixtures with compositions thought to be appropriate to
molecular clouds (d'Hendecourt \& Allamandola 1986;
Grim et al. 1989; Hudgins et al. 1993).
These spectra reveal no obvious absorption features near 3.25 $\mu$m.

We suggested earlier (Sellgren et al. 1994) that the 3.25 $\mu$m feature may
be due to absorption by aromatic hydrocarbons at low temperature,
based on a similarity in wavelength to the C--H stretch
of polycyclic aromatic hydrocarbons (PAHs)
isolated in neon matrices at a temperature of 4.2 K
(Joblin et al. 1994).
The aromatic C--H stretch wavelength
is a function of temperature,
increasing with increasing temperature
(Colangeli, Mennella, \& Bussoletti 1992; Joblin et al. 1994, 1995).
Aromatic hydrocarbons are a promising candidate for the
3.25 $\mu$m absorption feature, since
aromatic emission features at 3.3, 6.2, 7.7, 8.6, and 11.3 $\mu$m
have been observed
throughout the interstellar medium of our own and other galaxies.
Corresponding {\it absorption} features have been searched
for, but until now have not been definitely detected in molecular clouds.
The infrared emission features
have been attributed to a variety of
aromatic substances, including
hydrogenated amorphous carbon (HAC) grains
(Blanco, Bussoletti, \& Colangeli 1988; Ogmen \& Duley 1988),
PAHs
(L\'eger \& Puget 1984; Allamandola, Tielens, \& Barker 1985),
quenched carbonaceous composite (QCC) grains
(Sakata et al. 1987),
and other aromatic materials
(see Sellgren 1994 for a review of proposed identifications).

We compare in Table 1 the observed wavelength of the 3.25 $\mu$m
feature toward Mon R2/IRS-3 with the wavelengths of
several aromatic substances.
We list in Table 1 the measured wavelengths of
solid QCC (Sakata et al. 1990),
the 3.3 $\mu$m interstellar aromatic
emission feature (Tokunaga et al. 1991),
the PAH molecule coronene in the condensed phase
and the gas-phase (Flickinger, Wdowiak, \& G\'omez 1991),
solid HAC (Biener et al. 1994),
and the PAH molecules coronene and pyrene isolated in a neon matrix
(Joblin et al. 1994).

Joblin et al. (1995) have examined the temperature dependence of
the C--H stretch wavelength of gas-phase aromatic molecules in detail.
They state that the wavelength increases with increasing temperature due to
anharmonic coupling of the C--H stretch mode with excited longer
wavelength modes.
In Table 1 we also present
the predicted wavelengths for each aromatic material, when
shifted from the temperature at which
the measurement was made to a temperature of 80 K,
appropriate for the icy grains toward Mon R2/IRS-3 (Smith et al. 1989),
using Eq. 5 of Joblin et al. (1995) and the
assumption that the neon matrix does not introduce a wavelength
shift from the gas phase.
The temperature dependence of the aromatic C--H stretch wavelength
(Joblin et al. 1995)
was derived for gas-phase aromatic molecules,
and we caution that solid-phase aromatics, such as HAC or QCC,
may not follow the same relation.

In Figure 1, we compare the optical depth profile of the 3.25 $\mu$m
absorption feature and the
profile of the 3.3 $\mu$m aromatic
interstellar emission feature
in IRAS 21282+5050 (Nagata et al. 1988), after continuum subtraction
(Tokunaga et al. 1991), and after shifting
the center of the emission feature to
the predicted wavelength at 80 K (see Table 1).
The two feature profiles show reasonable agreement, although since
the width of each feature is probably dominated by
different processes, such agreement may be fortuitous.

The average of the observed feature wavelengths from
this paper and Sellgren et al. (1994) is 3.253 $\pm$ 0.004 $\mu$m,
which is shorter than the aromatic hydrocarbon wavelengths
in Table 1
by 0.004--0.032 $\mu$m.
The fact that the 3.25 $\mu$m absorption feature just barely overlaps
the short wavelength side of the range of cold aromatic hydrocarbon wavelengths
presents a problem,
since moving the aromatic C--H vibration to shorter wavelengths
(higher frequencies)
means strengthening the C--H bond, something that seems
difficult to achieve if the aromatic hydrocarbons
are immersed in an ice matrix of some sort.

Any identification of the 3.25 $\mu$m feature at this time
rests only on one absorption feature,
and the wavelength match with aromatic hydrocarbons is not exact.
A search for the longer wavelength features associated with
aromatic hydrocarbons would provide one test of this
identification.

If we assume that the 3.25 $\mu$m absorption feature is due to
aromatic hydrocarbons,
the column density of aromatic C--H bonds along the line of sight to
Mon R2/IRS-3 can be estimated.
Measurements of
aromatic hydrocarbons in
absorption are important because
estimates of the abundance of aromatic hydrocarbons
from the observed emission features
(Allamandola et al. 1989; Puget \& L\'eger 1989;
Joblin, L\'eger, \& Martin 1992)
are much less straightforward.

To estimate the column density of aromatic C--H bonds,
we use the relation, $N$ $\simeq$ $\tau
\Delta \nu$/$A$, where $\tau$ is the maximum optical depth of the 3.25
$\mu$m absorption feature, $\Delta
\nu$ is the feature FWHM in cm$^{-1}$, $A$ is the integrated
absorbance, and $N$ is the derived column density of molecular bonds
(Allamandola et al. 1992).
An average of the results of this paper and Sellgren et al. (1994) gives
$\tau$(3.25 $\mu$m) = 0.047 and $\Delta \nu$ =
66 cm$^{-1}$ for the 3.25 $\mu$m feature.
For the three aromatic molecules,
pyrene, coronene, and ovalene,
studied by Joblin et al. (1994),
the value of $A$ per
aromatic C--H bond for the 3.25 $\mu$m feature
was 0.7--1.4 $\times$ 10$^{-18}$ cm bond$^{-1}$ in the solid phase
and 2.1--4.1 $\times$ 10$^{-18}$ cm bond$^{-1}$ in the gas phase.
We average over all three molecules in both phases, to estimate
an average value of $A$ = 1.7 $\times$ 10$^{-18}$ cm
bond$^{-1}$.
We thus derive a column density of aromatic C--H bonds of $N$(C--H) $\sim$
1.8 $\times$ 10$^{18}$ bonds cm$^{-2}$ along the line-of-sight.

The abundance by number of aromatic C--H bonds, $X$(C--H), is the ratio
of the column density of aromatic C--H bonds divided by the total
hydrogen column density, $N_H$.  We estimate $N_H$ in two ways.
The silicate
optical depth, $\tau$(9.7 $\mu$m) = 4.3, observed toward Mon R2/IRS-3
(Willner et al. 1982)
implies $A_V$ = 80 mag
assuming A$_V$/$\tau$(9.7 $\mu$m) = 18.5 (Mathis 1990).
However, A$_V$/$\tau$(9.7 $\mu$m) is
observed to vary by a factor of two (Mathis 1990).
An independent estimate of $A_V$ comes from the 4.6 $\mu$m $^{13}$CO gas
absorption observed toward Mon R2/IRS-3 (Mitchell 1995), which
gives $N(^{13}$CO$)$ = 1.6 $\times$ 10$^{17}$ cm$^{-2}$.
If we assume
$A_V$/$N$($^ {13}$CO) = 4 $\times$ 10$^ {-16}$ cm$^2$ mag
(Dickman 1978),
then the $^{13}$CO gas column density implies
$A_V$ = 64 for Mon R2/IRS-3,
in good agreement with the value derived from the silicate feature.
We then convert our average value of $A_V$ = 72 to $N_H$ by
assuming $N_H$/A$_V$
= 1.9~$\times$~10$^{21}$~cm$^{-2}$~mag$^{-1}$ (Mathis 1990).
This implies
$N_H$ = 1.4 $\times$ 10$^{23}$ cm$^{-2}$ for Mon R2/IRS-3.
Again there is
some uncertainty in this because the value of $N_H$/A$_V$
measured in the diffuse interstellar medium
may not hold in molecular clouds.
Our derived value of $N_H$ implies that $X$(C--H) = 1.3 $\times$
10$^{-5}$ toward Mon R2/IRS-3.  For a solar abundance of carbon,
$X$(C)/$X$(H) =
3.6 $\times$ 10$^{-4}$ by number (Anders \& Grevesse 1989),
our estimate of $X$(C--H) implies that $\sim4$\% of the total carbon
along the line of sight toward Mon R2/IRS-3 is locked in aromatic
C--H bonds.

The total number of carbon atoms in aromatic hydrocarbons will be larger.
If the absorbing aromatic hydrocarbons have the same size distribution as the
emitting aromatic hydrocarbons, then we can use model results
for the interstellar emission features to estimate
the fraction, $f$,
of the number of carbon atoms in aromatic C--H bonds,
compared to the total number of aromatic carbon atoms.
The value of $f$ depends
on the aromatic hydrocarbon size, with a smaller value for larger
aromatic hydrocarbons.  D\'esert, Boulanger, \& Puget (1990) present a model of
interstellar dust, including size distributions for different grain components
and an analytic approximation for $f$ as a function of radius $a$ for
PAH molecules.
We have used their model,
with $a$ = 4--12 \AA\ for PAHs,
to calculate a size-averaged value for $f$ of 0.40.  The value
of $f$ for the absorbing aromatic hydrocarbons also
depends on the degree of dehydrogenation in
the interstellar medium, but aromatic hydrocarbons are predicted to be fully
hydrogenated in molecular clouds shielded from ultraviolet radiation
(Allamandola, Tielens, \& Barker 1989).  Thus the total amount of
carbon in aromatic hydrocarbons is
roughly a factor of $\sim$2.5 times higher than the
amount of carbon participating in aromatic C--H bonds.
If the 3.25 $\mu$m feature is due to absorbing aromatic hydrocarbons with
a size distribution
similar to that adopted by D\'esert et al. (1990)
for the emitting aromatic hydrocarbons
in the interstellar medium, this would
make the fraction of carbon in aromatic hydrocarbons $\sim$9\%.
If the absorbing aromatic C--H bonds are instead attached to larger
structures,
for instance if the aromatic absorption is due to hydrogen on the
surfaces of
large amorphous carbon grains while the aromatic emission is due to
small PAH molecules, then the
fraction of carbon in such structures would be much larger
than we estimate from the D\'esert et al. (1990) model.

Our estimate of the carbon abundance in aromatic hydrocarbons of $\sim$9\%
falls within the range of
previous estimates for the
aromatic hydrocarbon abundance,
which vary from 0.8\% to 18\%
of the total carbon abundance
(Lepp et al. 1988; Allamandola et al. 1989; Puget \& L\'eger 1989;
Joblin, L\'eger, \& Martin 1992).
Thus if the 3.25 $\mu$m feature is due
to aromatic hydrocarbons, we estimate that
a significant fraction of carbon remains in
aromatic hydrocarbons in molecular cloud dust.

If the 3.25 $\mu$m feature is due to, or contains contributions from,
non-aromatic species,
then the abundances of aromatic hydrocarbons along the line-of-sight derived
above become upper limits.  If it can be shown that {\it none} of the
feature is due to aromatic hydrocarbons,
then the abundance of carbon trapped in aromatic hydrocarbon
molecules may be much lower in molecular clouds than in photodissociation
regions or the diffuse interstellar medium.
Aggregation of aromatic hydrocarbon molecules into
larger graphitic-like structures is
one possible explanation.

The most pressing need is to detect the 3.25 $\mu$m feature in other
sources, both protostars and field stars behind molecular clouds.
Brooke et al. (1995) have recently
detected the 3.25 $\mu$m feature
toward the protostars NGC~7538/IRS-1 and S~140/IRS-1,
but observations
are needed over a wider range of physical conditions.
This will determine whether the feature arises in circumstellar
environments or in the surrounding molecular cloud, and constrain the
volatility of the absorber.

\vskip 24pt

\acknowledgments

We would like to thank Dolores Walther for assistance with these observations,
which were obtained during UKIRT Service Observing.
We also appreciate useful conversations with
Lou Allamandola, Christine Joblin, Scott Sandford,
and Alan Tokunaga.

\clearpage
\begin{center}
\begin{tabular}{llrll}
\multicolumn{5}{c}{{\bf Table 1: Possible 3.25 Micron Feature
Identifications}}\\[12pt]
&Measured&Measured&Predicted $\lambda$\\
Source&$\lambda$ ($\mu$m)&$T$ (K)&at 80 K ($\mu$m)&Ref.\\
\hline\hline\\
Mon R2/IRS-3&3.249 $\pm$ 0.004&80&3.249 $\pm$ 0.004&1\\
Mon R2/IRS-3&3.256 $\pm$ 0.003&80&3.256 $\pm$ 0.003&2\\
matrix-isolated coronene&3.257&4&3.257&3\\
gas-phase coronene&3.276&698&3.258&4\\
interstellar emission feature&3.289&1000&3.260&5\\
hydrogenated amorphous carbon&3.271&300&3.266&6\\
condensed coronene&3.290&788&3.268&4\\
matrix-isolated pyrene&3.268&4&3.269&3\\
quenched carbonaceous composite&3.289&300&3.285&7\\
\hline\\
\end{tabular}
\end{center}
\vskip 12pt\noindent
References---
(1) Sellgren et al. (1994);
(2) this paper;
(3) Joblin et al. (1994);
(4) Flickinger et al. (1991);
(5) Tokunaga et al. (1991);
(6) Biener et al. (1994);
(7) Sakata et al. (1990).

\vskip 12pt\noindent
Note:
The wavelength of these
aromatic substances at a
temperature of 80 K, appropriate for Mon R2/IRS-3 (Smith et al. 1989),
was predicted from the measured wavelength and the temperature
at which the wavelength was measured,
using the temperature-dependent wavelength shifts
measured by Joblin et al. (1995) for pyrene (for
pyrene) or coronene (for all other substances).
For the interstellar aromatic emission feature, we assumed a
particle temperature of $\sim$1000K
(Sellgren, Werner, \& Dinerstein 1983).

\clearpage
\centerline
{\bf Figure Captions}

\vskip 12pt

\noindent
{\bf Figure 1---}
New observations of the protostar Mon R2/IRS-3.
Gaps in the data near 3.30 $\mu$m and 3.32 $\mu$m are due to
Pfund $\delta$ in the standard star and strong telluric methane
absorption, respectively.
{\it Top}:
the 3.16--3.55~$\mu$m spectrum
({\it histogram})
with a resolution of 0.0033 $\mu$m
($\lambda$/$\Delta \lambda$~=~1000 at 3.25 $\mu$m).
The units are flux density ($F _ \lambda$)
in W cm$^{-2}$ $\mu$m$^{-1}$ vs. wavelength in microns.
A third-order polynomial
({\it solid curve})
was fit to the observations,
excluding 3.2--3.3~$\mu$m and 3.4--3.6~$\mu$m from the fit,
to determine the continuum.
{\it Middle}:
the 3.16--3.55~$\mu$m optical depth
({\it histogram}),
compared to the sum of two Gaussians ({\it solid curve}),
centered at 3.256~$\mu$m and 3.484~$\mu$m.
The central wavelengths, widths, and optical depths
of these two Gaussians were varied to
produce the best fit to the data.
{\it Bottom}:
the 3.16--3.55~$\mu$m optical depth
({\it histogram}),
compared to the profile of the aromatic
interstellar emission feature
({\it solid curve})
in IRAS 21282+5050 (Nagata et al. 1988), after continuum subtraction
(Tokunaga et al. 1991).
The emission feature profile was first shifted to bluer
wavelengths by 0.0294 $\mu$m to correct for temperature (see text and Table 1),
and then scaled by the ratio of the average 3.17--3.28 $\mu$m
optical depth of Mon R2/IRS-3 to the average 3.17--3.28 $\mu$m
feature profile of IRAS 21282+5050.

\pagestyle{empty}
\clearpage
\begin{figure}
\plotfiddle{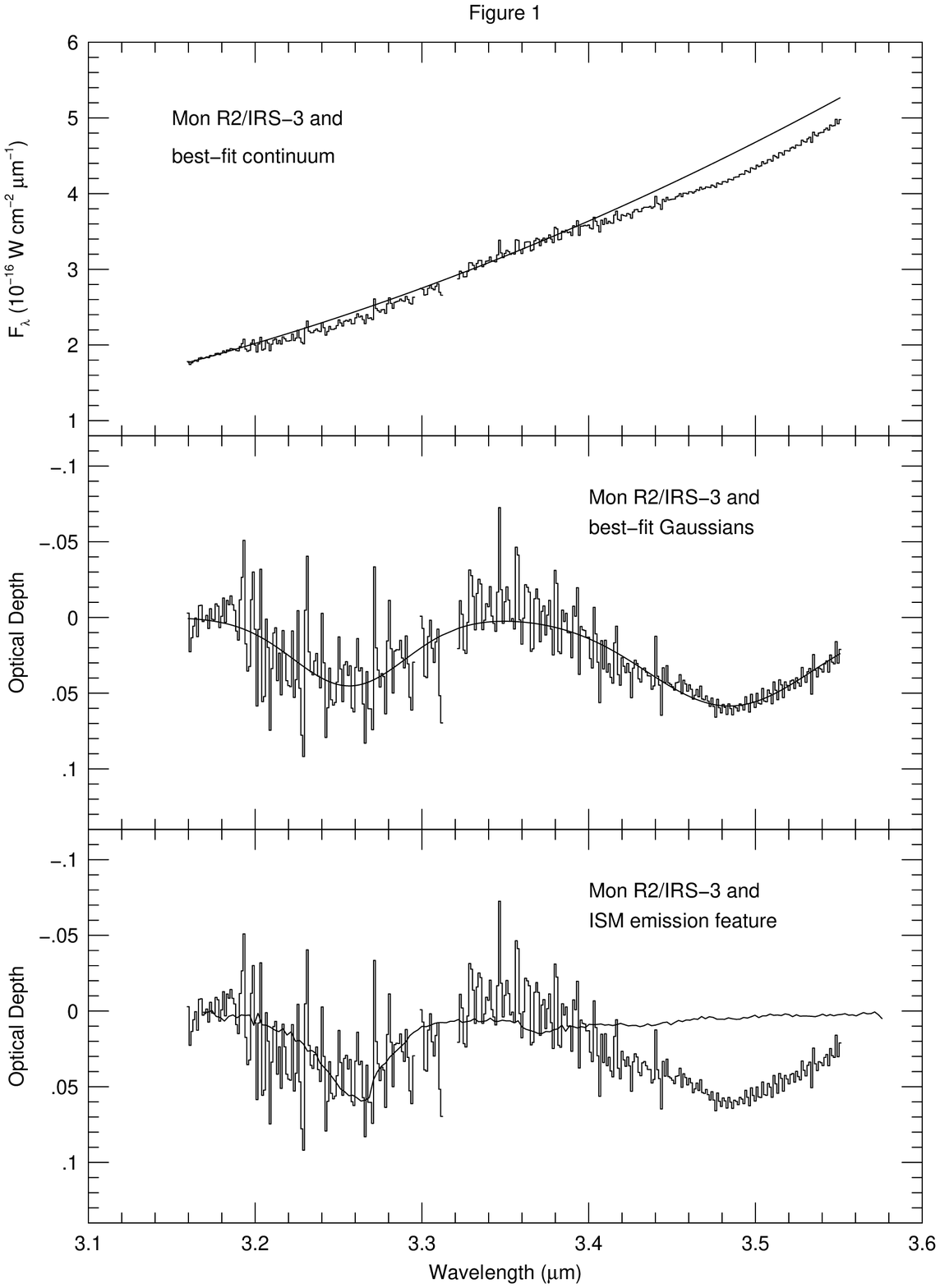}{9.0in}{0}{100}{100}{-324}{0}
\end{figure}

\end{document}